\documentclass[twocolumn,  superscriptaddress, pra]{revtex4}
\usepackage{amsmath}
\usepackage{amssymb}
\usepackage{graphicx}
\usepackage{booktabs}
\usepackage{subfigure}
\usepackage{esint}
\usepackage{tabularx}

\begin{document}

\title{Robust photon transmission in nonlinear parity-time-symmetric cavities%
}
\author{Ling-Pu Gong}
\affiliation{Department of Physics, Chongqing University, Chongqing 401330, China}
\author{Xing-Sen Chen}
\affiliation{Department of Physics, Chongqing University, Chongqing 401330, China}
\author{Yin Tan}
\affiliation{Department of Physics, Chongqing University, Chongqing 401330, China}
\author{Rui Zhang}
\affiliation{Department of Physics, Chongqing University, Chongqing 401330, China}
\author{Yu-Yu Zhang}
\email{yuyuzh@cqu.edu.cn}
\affiliation{Department of Physics, Chongqing University, Chongqing 401330, China}
\date{\today }

\begin{abstract}
We explore the photon transfer in the nonlinear parity-time-symmetry system
of two coupled cavities, which contains nonlinear gain and loss dependent on
the intracavity photons. Analytical solution to the steady state gives a
saturated gain, which satisfy the parity-time symmetry automatically. The
eigen-frequency self-adapts the nonlinear saturated gain to reach the
maximum efficiency in the steady state. We find that the saturated gain in
the weak coupling regime does not match the loss in the steady state,
exhibiting an appearance of a spontaneous
symmetry-breaking. The photon transmission
efficiency in the parity-time-symmetric regime is robust against the
variation of the coupling strength, which improves the results of the
conventional methods by tuning the frequency or the coupling strength to
maintain optimal efficiency. Our scheme provides an experimental platform
for realizing the robust photon transfer in cavities with nonlinear
parity-time symmetry.
\end{abstract}

\date{\today }
\maketitle

\section{Introduction}

Considerable progress has been made in parity-time (PT)- symmetric optics to
explore intriguing properties~\cite{makris08,klaiman,kottos,suchkov,lv15},
which has been investigated both theoretically~\cite%
{bender,klaiman08,jin10,hassan15} and experimentally~\cite{ruter,peng,sun14}%
. PT symmetric system, despite being non-Hermiticity, has been found novel
phenomenons, such as PT symmetry breaking~\cite{ruter,peng}, and
nonreciprocal reflectionless transmission~\cite{feng2013}. Even through the
Hermiticity of the quantum observables was never in doubt, a wide class of
non-Hermitian Hamiltonians with PT symmetry have motivated discussions on
several fronts in physics, including cavity optomechanics~\cite%
{jing,fan,peng12}, quantum field theories~\cite{bender04}, PT-symmetric
lattice~\cite{longhi,zhen15}, and open quantum system~\cite{rotter,dembowski}%
. A variety of optical structures provide a alternative platform for testing
various theoretical proposals on non-Hermitian PT-symmetric quantum mechanic~%
\cite{chang14}.

PT-symmetric systems are invariant under under simultaneous parity-flip and
time-flip operations~\cite{ganainy07,chong11}. In optical systems, PT
symmetry can be established by incorporating gain and loss in coupled
resonators. A linear PT-symmetric system with balanced gain and loss can
exhibit a real eigenvalue spectrum and present unusual properties~\cite%
{jin17,guo09}. If the gain-loss contrast exceeds a certain threshold, the PT
symmetry can be spontaneously broken and the spectrum is no longer entirely
real, exhibiting an exceptional point. When the gain induces large nonlinear
saturation under high pumping and no longer matches the loss, the linear PT
symmetry of the coupled cavities breaks down. Theoretically, the nonlinear
gain saturation can causes the system to reach a steady state that still
contains the PT symmetry characteristics~\cite{fan17,hassan,ge}. Can such
nonlinear gain saturation effects be harnessed to enhance the photon
transmission in coupled cavities? Photons transmission, such reflection and
transmission as a consequence of the nonlinear PT symmetry motivates us to
achieve an optimal photon transfer scheme.

We focus on how PT symmetry relates to the nonlinear gain and loss system
leads to a robust photon transfer in two coupled cavities. Different from
the linear PT-symmetric system with fixed value of gain, we consider a
nonlinear PT-symmetric system with a nonlinear gain depending on intracavity
photons. An analysis based on quantum Langevin equations is presented to
study the steady-state dynamics. It is observed that the eigenfrequency
self-adjusts to the nonlinear gain saturation, and the saturated gain in the
steady state automatically satisfies PT symmetry. As a consequence of the PT
symmetry with nonlinear gain saturation, the photons transfer efficiency
obtained is robust independent on the coupling strength, exhibiting an
improvement over a conventional transfer scheme, in which the input light
frequency is required to adjusted as the coupling strength to maintain
optimal efficiency.

\section{Conventional scheme}

To explore the photon transfer scheme in coupled cavities, we consider an
input cavity coupling to an output cavity. The Hamiltonian is described by%
\begin{equation}
H=\omega _{1}a_{1}^{\dagger }a_{1}+\omega _{2}a_{2}^{\dagger }a_{2}+\kappa
(a_{1}^{\dagger }a_{2}+H.c.),
\end{equation}%
where $a_{i}^{\dagger }$ ($a_{i}$) is the creation (annihilation) operator
of the cavity $i$ with the frequency $\omega _{i}$, $\kappa $ is the
coupling strength between two cavities.

\begin{figure}[tbp]
\includegraphics[trim=100 150 100 120,scale=0.3]{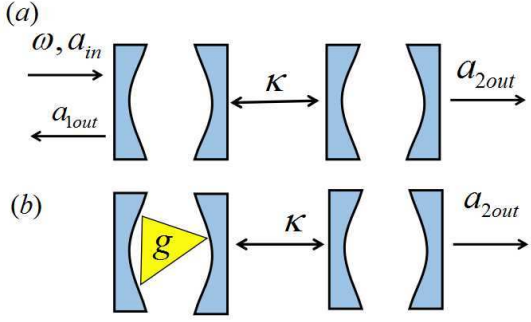}
\caption{(a)Conventional scheme. A driving at a frequency $\protect\omega $
interacts with the input cavity at a rate $\protect\gamma _{1}$. The input
cavity couples to the output cavity with the coupling strength $\protect%
\kappa $, which outputs photons at a rate $\protect\gamma _{2}$. The
intrinsic loss rate of two cavities are $\protect\gamma _{0}$. (b)
PT-symmetric scheme. We consider a nonlinear gain in the input cavity with
high pumping theoretically.}
\label{scheme}
\end{figure}
Fig.~\ref{scheme}(a) shows a conventional scheme of photon transfer in two
coupled cavities. A pump light field $a_{in}$ of a frequency $\omega $
transfers photons to the input cavity at a rate $\gamma _{1}$, and the
output cavity out-puts photons at a rate $\gamma _{2}$ due to the leakage of
photons. The intrinsic loss rate of each cavity is $\gamma _{0}$.

The dynamics of the photon transfer process undergoes dissipation due to
leakage of photons. It can be described by the quantum Langevin equations
for the Heisenberg operators $a_{1}$ and $a_{2}$,
\begin{eqnarray}
\dot{a}_{1}(t) &=&(-i\omega _{1}-\gamma _{0}-\gamma _{1})a_{1}(t)-i\kappa
a_{2}(t)+f_{1}(t), \\
\dot{a}_{2}(t) &=&(-i\omega _{2}-\gamma _{0}-\gamma _{2})a_{2}(t)-i\kappa
a_{1}(t)+f_{2}(t),
\end{eqnarray}%
where $f_{1}(t)=\sqrt{2\gamma _{1}}a_{in}(t)$ is the quantum Langevin force
originating from the input light field $a_{in}(t)$. For simplicity, the
force term $f_{2}(t)$ is neglected.

To solve above equations in frequency space, we employ a transforming
operator as $a(\omega )=\int_{-\infty }^{+\infty }e^{i\omega t}a(t)dt$ of $%
a(t)$. The input-output relations are given by
\begin{eqnarray}
a_{1out}(\omega ) &=&\sqrt{2\gamma _{1}}a_{1}(\omega )-a_{in}(\omega
)=R(\omega )a_{in}(\omega ), \\
a_{2out}(\omega ) &=&\sqrt{2\gamma _{2}}a_{2}(\omega )=T(\omega
)a_{in}(\omega ),
\end{eqnarray}%
where $a_{1out}$ and $a_{2out}$ are the reflected and output operators of
cavities, respectively. For an input light field at frequency $\omega $, the
transmission function $T(\omega )$ of the output cavity is obtained as%
\begin{equation}
T(\omega )=-\frac{i\kappa \sqrt{2\gamma _{2}}\sqrt{2\gamma _{1}}}{\kappa
^{2}+[i\Delta_1+\gamma _{0}+\gamma _{1}][i\Delta_2+\gamma _{0}+\gamma _{2}]},
\end{equation}%
and the response function $R(\omega )$ of the input cavity for the
reflection is%
\begin{equation}
R(\omega )=\frac{2\gamma _{1}}{i\Delta_1 +\gamma _{0}+\gamma _{1}+\kappa
^{2}/[i\Delta_2 +\gamma _{0}+\gamma _{2}]}-1,
\end{equation}%
where $\Delta_{1(2)} =-\omega +\omega _{1(2)}$. The corresponding photon
transfer efficiency from output cavity is given by
\begin{eqnarray}
\eta &=&\frac{\langle a_{2out}^{\dagger }a_{2out}\rangle }{\langle
a_{in}^{\dagger }a_{in}\rangle }=|T(\omega )|^{2},  \notag \\
\end{eqnarray}%
which, for a symmetric case ($\gamma _{1}=\gamma _{2}=\gamma_0 $ and $%
\omega_1=\omega_2=\omega_0$), becomes
\begin{equation}  \label{effi}
\eta =|\frac{2\kappa \gamma }{\kappa ^{2}+[i\Delta +2\gamma _{0}]^{2}}|^{2}.
\end{equation}%
with the detuning frequency $\Delta =-\omega +\omega _{0}$. In the
conventional photon transport process, it involves reflections $R(\omega)$,
which plays a role in the photon transport from the input cavity.

\begin{figure}[tbp]
\includegraphics[scale=0.65]{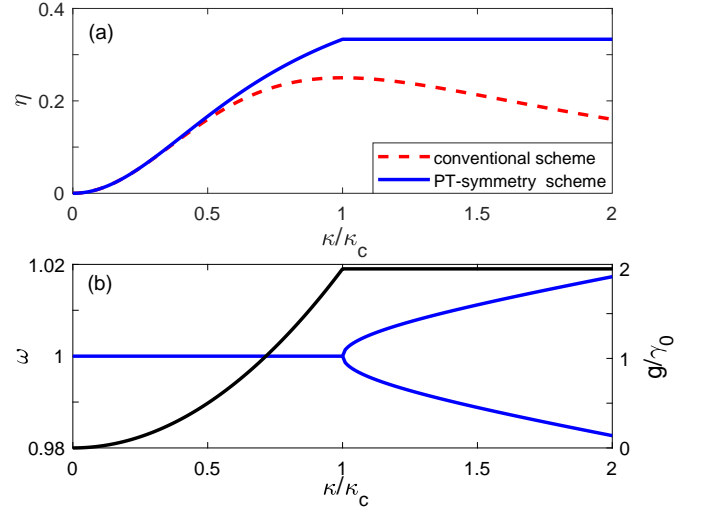}
\caption{(a)Photon transfer efficiency as a function of the scaled coupling
strength $\protect\kappa /\protect\kappa _{c}$ for a conventional scheme
(red dashed line) with the input frequency $\protect\omega=\protect\omega_0$
and a PT-symmetry scheme (blue solid line). (b)Steady-state mode frequency $%
\protect\omega $ and the saturated gain $g$ as a function of the scaled
coupling strength $\protect\kappa /\protect\kappa _{c}$. The parameters are $%
g_{0}=3f_{0}$, $\protect\gamma _{0}=0.005\protect\omega $, and $\protect%
\kappa _{c}=2\protect\gamma _{0}$. }
\label{efficiency}
\end{figure}

Fig.~\ref{efficiency}(a) shows the transfer efficiency of photons with a
resonance frequency $\omega =\omega _{0}$. It behaves non-monotonically
dependent on the coupling strength $\kappa $. The type of the conventional
photon transfer is not robust against coupling strength in the operating
conditions. To maximize the transfer efficiency $\eta $, it is required $%
\omega =\omega _{0}\pm \sqrt{\kappa ^{2}-4\gamma _{0}^{2}}$ for $\kappa
\geqslant 2\gamma _{0}$ by minimizing the module of the denominator in Eq.(%
\ref{effi}). One can tune the input frequency according to the coupling
strength $\kappa$ to achieve the optimal value of $\eta $.

\section{Nonlinear PT-symmetric scheme}

Different from the conventional scheme, the PT-symmetric scheme circumvent
photon reflections. The linear PT-symmetric system of two coupled resonators
with balanced gain and loss was previously investigated, in which a gain can
be induced by pumping erbium ions in the gain resonator\cite{chang14,jing}.
Here, we consider the lasing of the gain medium provides the nonlinear gain
theoretically, which depends on the intracavity photons~\cite{fan17,hassan}.
Fig.~\ref{scheme}(b) illustrates a nonlinear saturable gain element into the
input cavity for amplification of photons generation and a nonlinear loss in
the output cavity. The nonlinear gain rate $g$ and loss rate $\gamma $
depend on the intracavity average photons $\langle a_{1(2)}^{\dagger
}a_{1(2)}\rangle=|\alpha _{1(2)}|^{2}$, which are expressed as
\begin{equation}
g(\alpha _{1})=-\gamma _{0}+\frac{g_{0}}{1+|\alpha _{1}|^{2}},\quad \gamma
(\alpha _{2})=\gamma _{0}+\frac{f_{0}}{1+|\alpha _{2}|^{2}},
\end{equation}%
with the unsaturated gain and loss rates $g_{0}$ and $f_{0}$, respectively.

The Heisenberg operators $a_{1}$ and $a_{2}$ obey
\begin{equation}  \label{e1}
\dot{a}_{1}(t) =[-i\omega _{0}+g(\alpha _{1})]a_{1}(t)-i\kappa a_ {2}(t),
\end{equation}
\begin{equation}
\dot{a}_{2}(t) =[-i\omega _{0}-\gamma (\alpha _{2}) ]a_{2}(t)-i\kappa
a_{1}(t).  \label{e2}
\end{equation}
Since the nonlinear gain $g$ depends on the mean photons in the input
cavity, we explore the stable state with a saturable gain by solving the
above equations analytically.

\textit{PT-symmetric regime.--}A enhancement of the nonlinear gain and loss
generates large steady-state amplitudes in two cavities. We use the
mean-field description for the operators $\alpha_{i}=\langle
a_{i}\rangle=\alpha _{i0}e^{-i\omega t}$~\cite{lv16,lv}, where $\alpha _{i0}$
is the steady-state amplitude of the cavity mode. According to Eqs.(\ref{e1}%
) and (\ref{e2}), the steady-state amplitudes is given by the following
equations
\begin{equation}  \label{e3}
-i(\omega-\omega_0) \alpha _{10} =(-\gamma _{0}+g_{s})\alpha _{10}-i\kappa
\alpha _{20},
\end{equation}
\begin{equation}  \label{e4}
-i(\omega-\omega_0) \alpha _{20} =(-\gamma _{0}-f_{s})\alpha _{20}-i\kappa
\alpha _{10},
\end{equation}%
where the saturated gain and loss are $g_{s}=g_{0}/(1+|\alpha _{10}|^{2})$,
and $f_{s}=f_{0}/(1+|\alpha _{20}|^{2})$, respectively. The above equations
suggest the relation $\alpha _{02}=\rho e^{i\phi }\alpha _{01}$ with a phase
shift $\phi $ and the modal ratio $\rho \in {Re}^{+}$. It leads to the
equation for the eigen-frequencies
\begin{eqnarray}  \label{fre}
&&(\omega-\omega_0) ^{2}+i(\omega-\omega_0) (2\gamma _{0}-f_{s}-g_{s})
\notag \\
&&+(-\gamma _{0}+g_{s})(-\gamma _{0}-f_{s})-\kappa ^{2}=0.
\end{eqnarray}%
To obtain a real $\omega $, it is required $g_{s}/2\gamma _{0}
-f_{s}/2\gamma _{0}=1$. Then it is reasonable to give $g_{s}=2\gamma
_{0}\cosh ^{2}\eta $ and $f_{s}=2\gamma _{0}\sinh ^{2}\eta $ with a positive
real quantity $\eta $. It leads to the relations for the mean photons
\begin{eqnarray}
|\alpha _{10}|^{2} &=&\frac{g_{0}}{2\gamma _{0}\cosh ^{2}\eta }-1, \\
|\alpha _{20}|^{2} &=&\frac{f_{0}}{2\gamma _{0}\sinh ^{2}\eta }-1.
\end{eqnarray}%
Then the eigenfrequencies in Eq. (\ref{fre}) reduces to $(\omega-\omega_0)
^{2}=\kappa ^{2}-\gamma _{0}^{2}\cosh ^{2}(2\eta )$. By substituting into
Eqs. (\ref{e3}) and (\ref{e4}), one obtains $\rho =\pm 1$ and $\tanh \eta =%
\sqrt{f_{0}/g_{0}}$.

The saturated gain and loss are obtained as
\begin{eqnarray}\label{gama}
g_{s} =2\gamma _{0}\frac{g_{0}}{g_{0}-f_{0}}, f_{s} =2\gamma _{0}\frac{f_{0}%
}{g_{0}-f_{0}}.
\end{eqnarray}%
Therefore, the stable state is in the PT-symmetric phase with the balanced
gain and loss
\begin{equation}  \label{gain}
g_{\mathtt{PT}}(\alpha _{1})=\gamma_{\mathtt{PT}}(\alpha _{2})=\gamma_0\frac{%
g_0+f_0}{g_0-f_0}.
\end{equation}
The eigenfrequencies are obtained as%
\begin{equation}
(\omega-\omega_0) ^{2}=\kappa ^{2}-\gamma _{0}^{2}(\frac{g_{0}+f_{0}}{%
g_{0}-f_{0}})^{2},
\end{equation}%
which is real for $\kappa >k_{c}=\gamma _{0}(g_{0}+f_{0})/(g_{0}-f_{0})$. By
setting $g_{s}=3\gamma _{0}$ and $f_{s}=\gamma _{0}$ with $g_{0}=3f_{0}$,
the eigenfrequencies reduce into
\begin{equation}  \label{freq}
\omega =\omega_0\pm \sqrt{\kappa ^{2}-4\gamma _{0}^{2}},
\end{equation}
with the PT-symmetry $g_{\mathtt{PT}}=\gamma_{\mathtt{PT}} =2\gamma _{0}$.
The eigenfrequencies are consistent with the optimal frequency to maximize
the transfer efficiency in Eq. (\ref{effi}) in the conventional scheme. In
contrast to the conventional scheme with frequency tuning, the advantage of
the PT-symmetric scheme lies in the self-selected eigenfrequencies without
any active tunings.
\begin{figure}[tbp]
\includegraphics[scale=0.5]{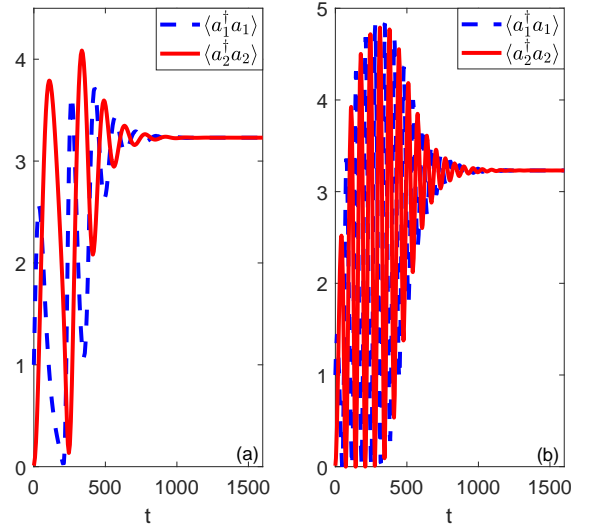}
\caption{Mean photons for each cavity in the PT-symmetry regime $\langle
a_{1}^{\dagger }a_{1}\rangle $ (blue dashed line) and $\langle
a_{2}^{\dagger }a_{2}\rangle $ (red solid line) obtained numerically as a
function of time $t$ for different coupling strength (a)$\protect\kappa/%
\protect\kappa_c =2$ and (b) $\protect\kappa/\protect\kappa_c =4 $. The
initial state is choose as $\protect\alpha _{1}(0)=1$ and $\protect\alpha %
_{2}(0)=0.1e^{i\protect\pi /2}$. The parameters are $\protect\kappa_c=1.1%
\protect\gamma_0$, $g_{0}=0.1\protect\omega $, $f_{0}=0.005\protect\omega $,
$\protect\gamma _{0}=0.0113\protect\omega $ and $\protect\omega _{0}=1$. }
\label{PT}
\end{figure}

Fig.~\ref{efficiency} (b) show the saturated gain and the self-adjusting
eigen-frequency $\omega $ in the steady state. We observe a bifurcation in
the real part of the frequency at the critical coupling strength $\kappa_{c}$%
, exhibiting a PT symmetry-breaking transition. In the strong coupling
regime $\kappa >k_{c}$, the system have two modes with eigenfrequencies $%
\omega =\omega_0\pm \sqrt{\kappa ^{2}-4\gamma _{0}^{2}}$. The corresponding
saturated gain in Eq. (\ref{gain}) is a constant independent on the coupling
strength, and can balance out the loss, $g_{\mathtt{PT}}=\gamma_{\mathtt{PT}}
$. It demonstrate that the system posses the PT symmetry in the strong
coupling regime $\kappa >k_{c}$.

The steady state in the PT-symmetric phase is given by
\begin{equation}
\left(
\begin{array}{c}
\alpha _{10} \\
\alpha _{20}%
\end{array}%
\right) =\sqrt{\frac{g_{0}-f_{0}-2\gamma }{2\gamma }}\left(
\begin{array}{c}
1 \\
e^{i\phi }%
\end{array}%
\right)
\end{equation}%
with $\sin \phi =\pm \gamma _{0}/\kappa $. One can easily obtain the mean
photons in the steady state for the input and output cavities, which have
the same value,
\begin{equation}  \label{PT state}
\frac{\langle a_{2}^{\dagger }a_{2}\rangle}{\langle a_{1}^{\dagger
}a_{1}\rangle}=\frac{|\alpha_{20}|^2}{|\alpha_{10}|^2}=1.
\end{equation}

Fig.~\ref{PT} shows the time evolution of the mean photons obtained by
solving Eqs.(\ref{e1}) and (\ref{e2}) numerically for different coupling
strength $\kappa $. It is observed that in the steady state $\langle
a_{2}^{\dagger }a_{2}\rangle $ equals to $\langle a_{1}^{\dagger
}a_{1}\rangle $, which is consistent with the analytical results in Eq. (\ref%
{PT state}). In the steady state, the mean photons $\langle
a_{1(2)}^{\dagger }a_{1(2)}\rangle$ for $\kappa/\kappa_c=2$ is the same as
that for $\kappa/\kappa_c=4$, which is independent on the coupling strength.

With the input and output relations $\hat{a}_{in}=\sqrt{2g_{s}}\hat{a}_{1},%
\hat{a}_{2out}=\sqrt{2f_{s}}\hat{a}_{2}$, the photon transfer efficiency is
obtained as
\begin{equation}
\eta =\frac{\langle \hat{a}_{2out}^{\dagger }\hat{a}_{2out}\rangle }{\langle
\hat{a}_{in}^{\dagger }\hat{a}_{in}\rangle }=\frac{f_{s}}{g_{s}},
\end{equation}
where $g_{s}$ and $f_{s}$ are the saturated gain and loss in Eq. (\ref{gama}).
So the efficiency is independent on the coupling 
strength $\kappa$ in the PT-symmetric regime.

Fig.~\ref{efficiency}(a) shows the transfer efficiency in the strong
coupling regime $\kappa >k_{c}$. The efficiency $\eta$ is robust against the
variation of the coupling strength $\kappa $. And $\eta$ is larger than that
obtained by the conventional transfer scheme, which varies dependent on the
coupling strength for a fixed input frequency. Since the eigen-frequency in
the non-linear PT-symmetric scheme can self-adjust to the variation of the
the coupling strength for an optimal transfer efficiency. It exhibits an
improvement over the conventional scheme due to the robust efficiency
induced by the saturated gain and self-adjusting frequency in the steady
state.

\textit{PT-broken regime.--} In the PT-broken regime for a weak coupling
strength $\kappa <\kappa _{c}$, the real eigenfrequency is given by $\omega
=\omega _{0}$ from Eq. (\ref{freq}). The mean-field approximation for the
stationary solutions is assumed to be $\alpha _{i}=\langle a_{i}\rangle
=\alpha _{i0}$. The equations of motion in the steady state can be given 
in the interacting representation
\begin{equation}
0=(-\gamma _{0}+g_{s})\alpha _{10}-i\kappa \alpha _{20},  \label{eq5}
\end{equation}%
\begin{equation}
0=(-\gamma _{0}-f_{s})\alpha _{20}-i\kappa \alpha _{10}.  \label{eq6}
\end{equation}%
By solving the above equations, one obtains the saturated gain
\begin{equation}
g_{s}=\gamma_0+\frac{\kappa ^{2}}{\gamma _{0}+f_{s}}.
\end{equation}
The corresponding gain and
loss in the stable state are obtained by setting $f_{s}=\gamma _{0}$
\begin{equation}
g_{\mathtt{b}}=-\gamma _{0}+g_{s}=\frac{\kappa ^{2}}{2\gamma _{0}},\gamma _{%
\mathtt{b}}=\gamma _{0}+f_{s}=2\gamma _{0},  \label{gpt}
\end{equation}%
Obviously, it exhibits the PT symmetry broken, because the saturated gain
does not match the loss $g_{\mathtt{b}}\neq \gamma _{\mathtt{b}}$.
Our results are consistent with the saturated gain in wireless power transfer
system with magnetic resonators~\cite{fan17}.

The ratio of mean photons in the input and output cavity is obtained
analytically as
\begin{equation}  \label{PTBR1}
\frac{\langle a_{2}^{\dagger }a_{2}\rangle}{\langle a_{1}^{\dagger
}a_{1}\rangle} =\frac{|\alpha_{20}|^2}{|\alpha_{10}|^2}=\frac{\kappa^2}{4\gamma_0^2},
\end{equation}
where $\alpha _{20}/\alpha _{10}=-i\kappa/(2\gamma_0)$ is obtained from Eq.(\ref{eq5}).
Obviously, the ratio depends on the coupling strength $\kappa$, which is
different from that in Eq.(\ref{PT state}) in the PT-symmetry regime.
Since the ration is smaller than $1$, giving $\kappa^2/4\gamma_0^2\leq1$,
it leads to the PT-broken regime $\kappa<\kappa_c$ with the critical
value $\kappa_c=2\gamma_0$.

One obtains the photon transfer efficiency
\begin{equation}
\eta =\frac{f_{s}\langle a_{2}^{\dagger }a_{2}\rangle }{g_{s}\langle
a_{1}^{\dagger }a_{1}\rangle }=\frac{\kappa^2}{2(\kappa^2+2\gamma _{0}^{2})}.
\end{equation}%
Fig.~\ref{efficiency}(a) shows the transfer efficiency in the PT-broken
regime for $\kappa<\kappa_c$.  $\eta$ increases as the coupling strength $%
\kappa$ increases. Fig.~\ref{efficiency}(b) shows that
the saturated gain in Eq. (\ref{gpt}) is proportional
to $\kappa^2$. It is different the photon transfer efficiency and the saturated gain
in the PT-symmetric regime for $\kappa>\kappa_c$, which is robust independent on
the coupling strength.

\section{Conclusion}

We have studied theoretically a nonlinear PT-symmetric scheme for photon
transfer in two coupled cavities by introducing a nonlinear gain and loss.
The gain induces nonlinearity dependent on the intracavity photons, which is
different from the linear PT-symmetric system with the balanced gain and
loss. We obtain the analytical solution to the steady state, which agree
well with the numerical ones. The nonlinear saturated gain and loss in the
steady state satisfy the PT symmetry automatically. And the eigen-frequency
self-adjusts to the nonlinear saturated gain, and exhibits a bifurcation at
the critical coupling strength. Consequence, we obtain the robust photon
transmission efficiency independent on the coupling strength, exhibiting an
improvement over the conventional scheme. In contrast to the conventional
scheme of the photon transfer involving the photon reflections, in which the
transfer efficiency depends on the coupling strength and input frequency
tuning, the nonlinear saturated gain guarantees the stable state with the
PT-symmetry and a self-selected frequency, which do not require active
tuning to maintain optimal efficiency. The proposal of the nonlinear
PT-symmetric system with nonlinear gain and loss provides a powerful
platform for investigating intriguing properties prior to those in linear
PT-symmetric systems, especially for compound photonic system.

\acknowledgments
We acknowledge useful discussions with Xin-You L\"{u}.
This work was supported by Fundamental Research Funds for the
Central Universities Grant No. 2020CQJQY-Z003.

\end{document}